\documentclass[twocolumn,pre,showpacs]{revtex4}

\usepackage{graphicx}
\usepackage{rotating}
\usepackage{amsmath}
\usepackage{amsfonts}
\usepackage{amssymb}
\usepackage{enumerate}
\usepackage{longtable}
\usepackage{dcolumn}
\usepackage{bm}
\usepackage{epsfig}
\begin{document}

\title{Opinion and community formation in coevolving networks}
\author{Gerardo I\~{n}iguez$^{1,2}$}
\author{J\'{a}nos Kert\'{e}sz$^{2,3}$}
\author{Kimmo K. Kaski$^2$ }
\author{R. A. Barrio$^{1,2}$}
\affiliation{$^1$Instituto de F{\'{\i}}sica, Universidad Nacional Aut\'onoma de M\'exico, Apartado Postal 20-364, 01000 M\'{e}xico Distrito Federal,
 Mexico,\email{barrio@fisica.unam.mx} }
\affiliation{$^2$Centre of Excellence in Computational Complex Systems Research, Department of Biomedical Engineering and Computational Science, Helsinki University of
 Technology, Espoo, Finland  \email{kaski@lce.hut.fi}}
\affiliation{$^3$Institute of Physics, Budapest University of Technology and
 Economics, Hungary \email{kertesz@phy.bme.hu}}
\date{\textrm{\today}}

\begin{abstract}
In human societies opinion formation is mediated by social interactions, consequently taking place 
on a network of relationships and at the same time influencing the structure of the network and its evolution. 
To investigate this coevolution of opinions and social interaction structure we develop a dynamic agent-based 
network model, by taking into account short range interactions like discussions between individuals, 
long range interactions like a sense for overall mood modulated by the attitudes of individuals, 
and external field corresponding to outside influence. Moreover, individual biases can be naturally 
taken into account. In addition the model includes the opinion dependent link-rewiring scheme to 
describe network topology coevolution with a slower time scale than that of the opinion formation. 
With this model comprehensive numerical simulations and mean field calculations have been carried 
out and they show the importance of the separation between fast and slow time scales resulting in the 
network to organize as well-connected small communities of agents with the same opinion.
\end{abstract}
\maketitle


\section{Introduction}
\label{intro}
The network approach has contributed significantly to our understanding of
the structure, function, and response of various complex systems, from genetic
transcriptions to human societies~\cite{RefWorks:133,Caldarelli}. In the case
of human social systems, this  approach was introduced by social scientists and
they established important concepts and tools to study 
them~\cite{granovetter2001eaa,wasserman1994sna}. According to the social network 
paradigm {\it social life consists of the flow and exchange of norms, values,
ideas, and other social and cultural resources channeled through a network}
\cite{white1976ssm}. Hence the networks serve as substrates for
various collective social phenomena such as diffusion and spreading processes
(of e.g. news and epidemics), opinion formation, language evolution, etc.

Until recently the studied empirical data sets of social systems remained rather limited since
the basic source of data were questionnaires, thus the focus had been on small 
scale properties. However, the most recent development in information-communication 
technology has opened the possibility to collect much larger data sets from Internet, emails,
phone records, etc.
\cite{RefWorks:127,albert-1999-401,ebel2002sft,eckmann2004edc,dodds2003ess,RefWorks:129}. 
While the scope of these records is narrow as compared to detailed questionnaires, 
their huge amount and objective quantifiability  enable us to study problems, which were
impossible to treat before, including the investigation of the structure and 
dynamics of entire populations \cite{sciencenature}. Much has been learned from 
these studies such as the broad distributions of network characteristics, the small
world properties, the modular organization of the social network, and the
relationship between the network topology and the intensity of the ties in the net.

One of the key problems in studying society and social dynamics falls under the term of
``opinion formation", which deals with a (measurable) response of the society to an issue, 
such as an answer to a political question or the acceptance of innovation. This problem
has attracted a lot of interest and a number of models of competing options have been 
introduced to study it~\cite{socdynam}. The simplest one is the so-called voter model~\cite{voter}, which 
has a binary opinion variable with the opinion alignment proceeding by a random 
choice of neighbors. Other discrete opinion formation models include those by 
Weidlich~\cite{weidlich1991} and Sznajd~\cite{sznajd}, where more than just a pair 
of spins is associated with the decision making procedure. These models are 
reminiscent to the Ising model and in the noiseless ($T=0$) case they usually approach 
an asymptotic state of consensus (ferromagnetic state). 

If these models are studied on complex networks instead of regular 
lattices~\cite{KZ2002, BDK2002, soodredner2005}, 
the temporal dynamics of the phase transition from one type of consensus to the 
other one may change significantly. Introducing noise or ``social temperature" makes 
the analogy with physics even closer and leads to a number of new phenomena from 
paramagnetic phase to stochastic resonance. In addition the introduction of an external field 
as the carrier of the influence of mass media is a natural and widely considered generalization
(see, e.g., \cite{KZ2002, HKS2001}). Also systems with more than two possible opinions 
have been considered \cite{VKR203}, which have found natural extension in models where the 
opinion of individuals is represented by a continuous variable \cite{HK2002,DNAW2000,SF2005}. 
By introducing a parameter of bounded confidence or tolerance that describes the situation where 
opinions close enough approach each other, the asymptotic state can be quite rich in allowing the 
coexistence of a number of different opinions. We also note that opinion dynamics is closely 
related to problems of competing cultures \cite{Axelrod1997} or languages \cite{Language}.

Opinion formation in a human society is mediated by social interactions between individuals, 
consequently it takes place on a network of relationships and at the same time it influences the 
network itself. Such behavior was analyzed in a seminal paper by Holme and Newman \cite{coevolution}.
Recently this problem of coevolving opinion formation networks was
considered for discrete \cite{NKB2008,KB2008} and continuous \cite{KBJPA2008} opinion dynamics. 
In these the separation of the time scales was controlled by the relative frequency of the 
opinion updating and rewiring processes, which is an important parameter of the
problem. It was found that the adjustment of the network has a major effect both on the 
opinion dynamics and on the resulting topology and that minor changes in the updating 
rules have dramatic consequences. In the case of discrete models it turned out that the
self-adaptation of the network resulted in the symmetry breaking of the 
asymptotic state with respect to the direction of the information transfer. In the case 
of continuous opinion formation dynamics rewiring tends on one hand to hinder the consensus
formation for large tolerance since it breaks the clusters. On the other hand there is a
tendency to decrease the number of different clusters for small tolerance. In all these 
investigations agents were considered to be uniform.

Here we also consider the continuous opinion formation in a coevolving network but with the 
addition that the individuals need not to be uniform. In particular we examine a situation in which 
these non-identical individuals form their opinions in information-transferring interactions with others. 
This could happen directly through discussions between individuals constituting the network structure or 
indirectly by sensing the overall mood or opinion of all the other individuals, the effect of which would 
depend on their personal attitude towards the overall opinion. We expect that in general the time 
scale for detectable changes in the network structure is considerably slower than the time scale 
for the direct social exchanges to take place. In order to describe this situation we have developed 
a dynamic network model, where we take into account short range interactions for direct discussions 
between pairs of individuals, long range interactions for sensing the overall opinion modulated by the 
attitude of an individual, and external field for outside influence. The opinion formation dynamics is 
assumed to take place with a fast time scale well separated from the slow time scale network topology 
coevolution included by opinion dependent link-rewiring processes of the kind introduced as the 
basic mechanisms of network sociology \cite{kossinets:2006,kumpula:prl}. Note that unlike the model 
introduced in \cite{KBJPA2008} our model does not contain any parameter for bounded confidence, 
which could alter the time scale of network coevolution, as we discuss later.

This paper is organized such that next we describe our model, followed by presenting the 
numerical results and then by finding analytical solutions to the dynamical equation with 
mean field approximation. Finally we discuss the results and draw conclusions.

\section{Description of the model}
\label{Model}

Let us consider an opinion formation in a network of a fixed number of individuals or 
agents ($N$) to whom a simple question is posed. We assume that the agents are unbiased but 
they have initial opinions concerning the issue of the question. The process of opinion formation 
is started by letting the agents to discuss or exchange their views with their acquaintances 
or friends for some time. After this some discussions are interrupted at regular intervals to let 
discussions with new agents to start. At any time an agent may decide to fix its opinion to either total 
agreement or disagreement, after which the decision is considered irrevocable and no longer
modifiable, although the agent continues to discuss and influence others. The whole process stops 
when everyone has balloted. It should be noted that here the aim is not to reach a consensus of 
opinion, but to give every agent a chance to use its social network to form a point of view 
through exchange of information. 

This kind of a social system can be described by a network where the nodes correspond to the agents 
and the edges to the links or interactions between them. Each individual agent $i$ is described with 
the time dependent state variable $x_i \in {[-1,1]}$ measuring the agent's instantaneous 
inclination towards the posed question. The agent's interactions with other agents are 
described with the adjacency matrix elements $A_{ij} \in {\lbrace{1,0}\rbrace}$ that represent the 
presence or absence of discussions between the pair of agents related to the question at hand. 
Thus the initial topology of the network defined by the adjacency matrix $\mathbf{A}$ changes when 
the agents are allowed to stop discussions. 

As for the evolution dynamics of this system we assume that the state variable $x_i$ of node $i$ 
depends on the links of the network, and likewise the links of the network change according to the values 
of the state variable of each node. These mutually dependent processes can be described in general 
with a dynamical equation of the following form
\begin{equation}
\label{eq:1}
\frac{dx_i}{dt} = \frac{\partial x_i}{\partial t} + \sum_j \hat{O}(x_i,x_j,g) A_{ij},
\end{equation}
which is based on the assumption of two separate time scales. Here we assume that many discussions 
take place before a change in the network structure occurs, that is, the time scale for each discussion or 
exchange of information (``transaction'') is $dt$, while the time scale for a change of connections in the 
network (``generation") is $T = gdt$. The quantity $g$ defines the number of transactions per generation 
and describes the separation between the fast transaction and slow generation time scales. 
In this equation $\hat{O}$ stands for an operator that changes the entries and/or the size of the adjacency 
matrix. The actions of the operator $\hat{O}$ are necessarily discrete and limited to few processes. In general 
there are only four basic ways to modify a simple network: either one deletes/creates links or deletes/creates 
nodes. Then a common operation of rewiring can be regarded as a composed operator with one 
deletion followed by one creation of a link. An important point in modeling a particular system is that the 
operator $\hat{O}$ should contain rules for link deletion or creation reflecting all the additional information 
about the system. 

For large $g$ or $T$ the effect of discreteness of network evolution events on the agent's state 
variable $x_i$ can be treated separately and the dynamics between
such updates is essentially continuous. Then Eq.~(\ref{eq:1}) 
can be approximated as follows 
\begin{equation}
\label{eq:2}
\frac{\partial{x_i}}{\partial{t}} = \alpha_i f_0(\lbrace{x_j}\rbrace) + f_1(\lbrace{x_j}\rbrace) x_i + h_i,
\end{equation}
where the first term on the right hand side represents the combined effect of an agent $i$ sensing the 
overall opinion of all the other agents ($f_0(\lbrace{x_j}\rbrace)$) but modulated by the agent's 
own attitude towards overall or public opinion. We denote this personal attitude by $\alpha_i$ and assume it 
being random and uniformly distributed between -1 and 1, where the former corresponds to completely 
opposing and the latter to completely agreeing attitude of the agent $i$ towards the overall opinion. 
The second term on the right hand side represents the direct discussions of the agent $i$ with the 
agents $j$ it is linked to. Note that the agents could be different also in their 
attitudes towards direct discussions, which would introduce another attitude parameter in the second term. 
Instead $\alpha_i$ should be considered as the relative attitude parameter. The third term on the 
right hand side, $h_i$, is an external field representing the personal bias towards either opinion (-1 or +1) 
due to e.g. mass media (newspapers, TV, radio). In Eq.~(\ref{eq:2}) $f_0$ and $f_1$ standing 
for the dynamic long and short range interaction terms are defined as follows
\begin{equation}
\label{eq:3}
f_0 = \sum_{\ell=2}^{\ell_{max}} \frac{1}{\ell} \sum_{j \in m_{\ell}(i)} x_j,
\end{equation}
and
\begin{equation}
\label{eq:4}
f_1 = \text{sign}(x_i) \sum_{j \in m_1(i)} x_j,
\end{equation}
where $m_{\ell}(i)$ means the set of nodes that are $\ell$ steps away from node $i$ (or $\ell^{th}$
neighbors of $i$) and $\ell_{max}$ is the number of steps needed to reach the most distant neighbors
of $i$ (or the maximum range of interactions for $i$).

Note that Eq.~(\ref{eq:2}) allows changes of sign in the state variable but also its exponential decay 
or growth such that $|x_i|$ could eventually become larger than one, which carries no meaning in our model.  
Thus we need first to detect the totally convinced agents and stop their dynamics, which can simply be done 
by a line of computer program ``if $\text{abs}(x_i) \ge 1$, then $x_i = \text{sign}(x_i)$", that is, these
agents cannot modify their state in subsequent times, but they are still linked to the network and taken
into account in the dynamical evolution of the undecided agents. 

In order to consider the other dynamical process (and the corresponding time scale) contained in
Eq.~(\ref{eq:1}), namely network connection topology changes at generation time intervals $T$,
we have here adopted the scheme of rewiring. There are basically two possibilities, either
\textit{global} or \textit{local} rewiring, which we will perform with probability $y$  and $1-y$,
respectively. These rewiring schemes are rooted to the fundamental link formation mechanisms of network
sociology proposed by Kossinet and Watts \cite{kossinets:2006}: \textit{focal closure} independently of
distance thus being global and \textit{triadic closure} between close network neighbors thus being
local in nature. In both these schemes, first the agent $i$ can choose to cut an existing link with an
agent $j$, i.e. end a discussion if their opinions are incompatible. In order to perform this process,
the quantity
\begin{equation}
\label{eq:5}
p_{ij} = A_{ij} \frac{|x_i - x_j|}{2}
\end{equation}
is calculated and all the links are put in ranking order. Then links with larger weights $p_{ij}$ are 
deleted first since they correspond to divergence of opinion. After this link deletion step follows the link 
creation step using either local or global rewiring scheme.

In the local rewiring scheme, an agent $i$ can create a link with a second neighbor by starting a discussion 
with the ``friend of a friend" if this new link can help the agent in reaching a state of total conviction
($|x_i| = 1$). In order to determine this we calculate the quantity
\begin{equation}
\label{eq:6}
q_{ij} = \left(1 - A_{ij}\right) \Theta \left[(A^2)_{ij}\right] \frac{|x_i + x_j|}{2},
\end{equation}
where the first two factors including the adjacency matrix (with $\Theta [x]$ being the discrete step
function $\Theta[x] = \sum_{k=1}^{N-2} \delta_{k,x}$), test the existence of a link to the second neighbor
and the third factor $|x_i + x_j|/2$ is the measure of similarity of opinions between the agents. Then all
the potential local rewiring links are put to ranking order, of which the links with larger weights $q_{ij}$
are created first. 

In the global rewiring scheme, an agent $i$ can create a link with further neighbors ($\ell > 2$) 
provided their opinions are similar. In order to determine this we follow a similar procedure as above by 
calculating instead the following quantity
\begin{equation}
\label{eq:7}
r_{ij} = \left(1 - A_{ij}\right) \left(1 - \Theta \left[(A^2)_{ij}\right]\right) \left(1 - \frac{|x_i - x_j|}{2}\right),
\end{equation}
where once again the first two factors test the far link existence and the third the similarity of opinions
between agents, followed by putting the potential links to ranking order and creating links with larger
weights first.

The rewiring of \textit{each} node $i$ is done in such a fashion that the number of links deleted is equal 
to the number of links created. Therefore, the total number of links should be conserved. This is clearly the 
case when the rewiring is performed sequentially, but in case of parallel rewiring (as we do) it could happen 
that if both agents $i$ and $j$ are performing rewiring simultaneously, their mutual actions could lead to the 
net creation or deletion of a link. Then the total number of links would not be conserved.

Next we will present the numerical results of our model followed by the analytical mean field theory treatment 
of the dynamical equation.

\section{Numerical results}
\label{nc}

For computer simulations we have first initialized our model system to a random network configuration of $N$ 
nodes and average degree $\left<k_0\right>$. This is done in the beginning of each simulation run to secure 
a different random configuration for the initial network and good statistics for the averaged quantities of
interest. As other initial conditions we chose the fixed constant $\alpha_i$ for the personal attitude and
the state variable $x_i(0)$ of agent $i$ randomly from a uniform distribution between -1 and 1 and from a
Gaussian distribution with unit standard deviation, respectively. In addition we chose all the external bias
field terms to be $h_i = 0$. 

With these initial choices we are left with only two parameters to vary, namely $y$ and $g$, upon 
which the average properties of the final network should depend and scale with the initial conditions 
$\left<k_0\right>$ and $N$. For the sake of simplicity we here set $y = 0$ in all our numerical 
simulations, which means that we take into account only the local rewiring to study the effect of the 
parameter $g$ alone.

Since our model includes fast transaction dynamics and slower generation dynamics for network rewiring 
the simulations have been carried out by using the following two-step process. In the first step the
dynamics of transactions described with Eq.~(\ref{eq:2}) is realized by numerical integration using a
simple Euler method, in which the time step was set to $dt = 10^{-4}$ as found to guarantee the stability
and reliability of the numerical calculations. By keeping the parameters fixed the system is driven until
the specified time ($gdt$) or $g$ time steps to then do the second step, namely the network rewiring with
the procedure described in the previous section. This two-step process is iterated until the system
reaches its final state, where no more changes in the $x_i$'s and in the network connections take place.
We have found that the way the system approaches its final state consists of two regimes, which can be
analyzed by controlling the number of changed links per cycle. The first regime shows a rapid roughly
exponential decay where both the number of agents that change opinions and the number of rewired links is
high. This fast decay crosses over to a very slow regime after most of the agents have reached either
opinion $x = 1$ or $-1$. During this slow decay, when frustrated links try to get optimal positions and the
remaining few undecided agents converge their opinions to $x = 1$ or $-1$, the network structure is found
not to change significantly. These few ``non-conformists'' that take extremely long time to get fully
convinced may serve as kind of nucleation centers if an external field (media effect) is switched on,
which is an issue we will investigate in the future.

From the simulation results we calculated the following averaged single site properties: the degree
$\left<k\right>$, the shortest path $\left<L\right>$, the average clustering coefficient $\left<C\right>$,
the mean number of second neighbors $\left<n^{(2)}\right>$, and the average cluster size or susceptibility
$\left<s\right> = \sum_s n_s s^2 / \sum_s n_s s$ that is the second moment of the number of $s$-size
clusters, see e.g.~\cite{Stauffer}. According to percolation theory the sums run over all the cluster sizes
excluding the giant component, as identified in finite samples with the largest cluster.

\begin{figure}[h!]
\begin{center}
\epsfxsize = \columnwidth \epsffile{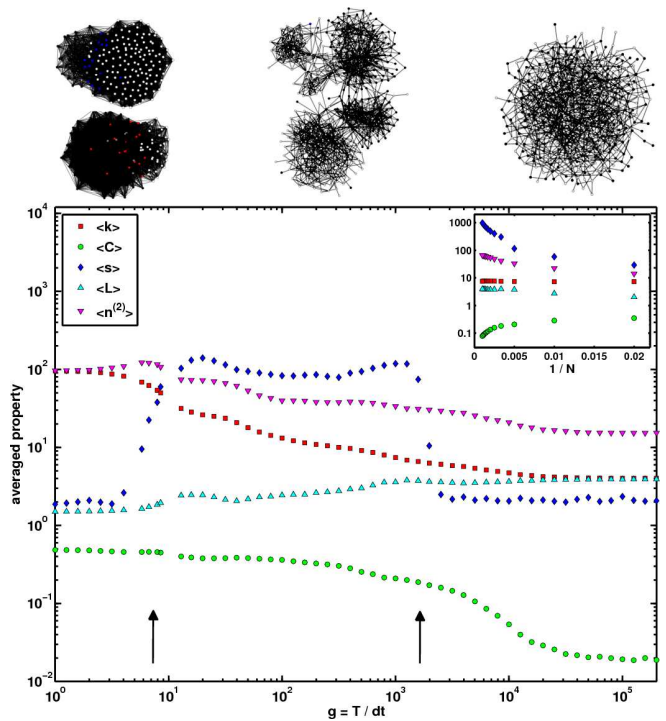}
\caption{Average degree $\left<k\right>$, average susceptibility $\left<s\right>$, average number 
of second neighbors $\left<n^{(2)}\right>$, average shortest path $\left<L\right>$, and average 
clustering coefficient $\left<C\right>$ for a network of size $N = 200$ and averages taken over 100 runs.
The arrows indicate the $g$ values where the susceptibility changes drastically. The inset shows such quantities
as a function of system size $N$ for $g = 1000$ and averaged over 1000 realizations. The graphs at the top are 
examples of final network configurations for $N = 400$ and $g = 5,10^3,10^5$, which show three different types
of configurations emerging. White circles mean agents with $x = 1$ and black those with $x = -1$.}
\label{fig:1}
\end{center}
\end{figure}

Fig.~\ref{fig:1} shows the various properties of the final network as a function of $g$, averaged 
over 100 realizations for the initial average degree $\left<k_0\right> = 4$ and for the system size 
$N = 200$.
Here one can visually distinguish three regions of behaviour.
For small $g \lesssim 10$ the dynamics of the state variable is practically irrelevant and accordingly,
the rewiring of the network is random and favours the formation of triangles. The final network 
appears divided into two large cliques of about $N/2$ nodes with almost no connections between them,
as expected from models without dynamics. 

On the other hand, if $g$ is larger than the mean number of iterations (i.e. $\gtrsim 10^4$) needed 
for the state variable to reach either of its extreme values at all sites, no rewiring of the network
is seen to occur. Thus the resulting network shows the properties of an Erd\"os-R\~enyi graph, i.e. 
$\left<k\right> = N \left<C\right>$, and $\left<L\right> = \ln N / \ln \left<k\right>$. Furthermore,
since the initial network is random no apparent clustering of the two opposite opinions is seen to develop. 

\begin{figure}[h]
\begin{center}
\epsfxsize = \columnwidth \epsffile{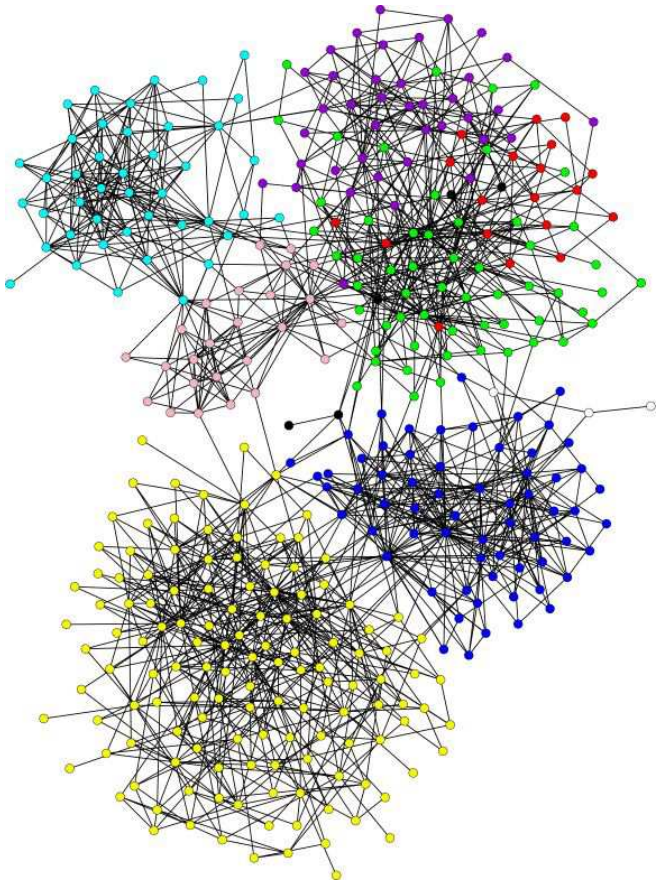} 
\caption{Communities found by the local fitness algorithm in the network configuration shown in the
middle top of Fig.~\ref{fig:1} for a system with $N = 400$, $\left<k_0\right> = 4$ and $g = 1000$. The
communities are distinguished by arbitrary coloring, regardless of the value of variable $x$ or opinion.}
\label{fig:2}
\end{center}
\end{figure}

For the intermediate $g$ values both dynamical processes, slow and fast, are at play and the net result is
that the final configuration of the network shows a segregation of several well-connected communities of
roughly similar size and of uniform opinion, as the ones shown in the middle top of Fig.~\ref{fig:1}. Such
result makes this model of opinion formation suitable to describe the complex community structure often found
in human societies (see, e.g., \cite{RefWorks:129,RefWorks:130}). It is relevant to mention that Mandr\`{a}
{\it et al.} \cite{mandra} have studied rewiring in the Ising-Glauber dynamics with tunable thresholds and
identified the emergent domains of spins with communities.

Such an emergence of medium size communities is clearly seen in the susceptibility $\left<s\right>$, which
peaks strongly (by two orders of magnitude) for the range of $g$ from about 10 to more than $10^3$. It should
be noted here that computationally the algorithm to solve Eq.~(\ref{eq:2}) in a rewiring network goes
roughly as $O(N^3)$, which makes the simulation of systems with more than 1000 nodes very time-consuming.
However, as it is seen in the inset of Fig.~\ref{fig:1}, the system size dependence of all quantities is
monotonic and can thus be extrapolated to the thermodynamic limit. Hence a subject of further study would
be to investigate the nature of the transitions between the observed morphological phases. In addition, we
would like to point out that all the other quantities show gradual behavior as a function of $g$. In the
models of opinion formation without network dynamics, the resulting network is either a single cluster of
consensus or two clusters of about the same size with opposite opinions. (As stated before we do not have
bounded confidence in our model, which could be another source of the creation of many communities
\cite{KB2008}.)

Note that to produce the graphs in Fig.~\ref{fig:1} we have used a visualization software~\cite{himmeli}
that tends to group agents according to their topological properties, and consequently communities are 
distinguishable by the eye. It is worth pointing out that within each community all its agents share 
the same opinion. In most previous studies the similarity of attributes between agents has not been considered 
to define communities (see for instance the recent comprehensive review by Fortunato \cite{fortunato}).

In order to find out whether the communities formed by agents of the same opinion match the ones formed by
topological similarities, we have used the community detection algorithm based on the local fitness concept 
described by Lancichinetti {\it et al.} \cite{RefWorks:104}. In Fig.~\ref{fig:2} we show the results for the
network configuration at the middle top of Fig.~\ref{fig:1} by using Eq.~(1) of Ref.~\cite{RefWorks:104},
with the resolution parameter set to $0.9$. Observe that in Fig.~\ref{fig:1} the color coding for the opinions
allows the eye to distinguish four well-defined communities, while in Fig.~\ref{fig:2} the algorithm based
on the local fitness separates the network into {\em seven} communities. Note that although there is in general
good correspondence between these two partitions, the community detection algorithm reveals sub-structure that
the eye does not capture.

The key point is to investigate what are the circumstances that drive the system to form many communities of
approximately the same size. Here the role of the personal attitude parameter $\alpha_i$ is very important,
since a network with agents having all positive $\alpha$'s  should reach consensus quite rapidly without
community structure forming, while a network with only negative $\alpha$'s should separate into two clusters
of different opinions. However, in case of $\alpha$'s being randomly distributed (as is assumed in this study)
frustration is introduced to the system, making the network structured.

It is instructive to look at the distribution of the $\alpha$'s and how it relates to the cluster structure.
For small values of $g$, where the rewiring process is very rapid and only two clusters eventually develop,
the attitude parameters have a minor role, so the $\alpha$ distributions in the two clusters are broad and 
similar. However, for the intermediate $g$ values the situation is different. Here the smaller clusters have
a rather narrow distribution with mostly negative $\alpha$ values, while the distributions for larger clusters
are broad and shifted toward positive $\alpha$ values. Naturally, the agents with negative $\alpha$'s do not
feel comfortable in a large homogenous cluster, thus they tend to build smaller communities. It is interesting
that the personal attitude of an agent has such an important effect on the clustering properties of the system
in spite of the fact that it is not defined as a bias towards a certain opinion but rather as an agent's
ability or intention to adjust or not to any kind of opinion of the other agents.

\begin{figure}[h!]
\begin{center}
\epsfxsize = \columnwidth \epsffile{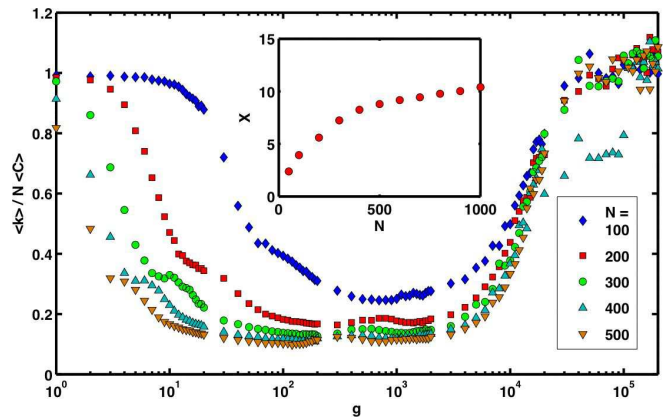}\\
\caption{The ratio $\left<k\right> / (N \left< C\right>)$ as a function of $g$, for various values of $N$.
Each point corresponds to an average over 100 realizations. The inset shows the number of communities $X$
as a function of $N$ for $g = 1000$ and averaged over 1000 realizations.}
\label{fig:3}
\end{center}
\end{figure}

The high connectivity and clustering coefficient of the final network configuration is the result of the 
appearance of communities: if there are $X$ communities of size $n$, then $N = Xn$ and if the clusters
continue to be random, then $\left<k\right> / \left<C\right> = n$. Therefore, the region of values of $g$
for which $\left<k\right> / \left<C\right>$ is constant is where there are communities of size $N/X$. In
Fig.~\ref{fig:3} we show this ratio as a function of $g$ for networks of different size. Notice that the
region where communities are formed is wider for larger networks and that as the network size is increased
the number of clusters $X$ increases as well. In the inset we show the value of $X$ for networks of various
sizes, averaged over 1000 realizations and keeping $g = 1000$ constant. It is worth noticing that the mean
field prediction for a network of size $N = 400$ is that the system contains about eight communities on
average, which agrees well with the seven communities found by the local fitness approach used in
Fig.~\ref{fig:2}.

In addition we have calculated the average neighborhood degree as a function of node degree and found that
our final networks show \textit {assortative mixing}, i.e. high degree nodes are connected to other high
degree nodes, as is common in other social networks. As a kind of hypothetical exercise we have found that
if we modify Eq.~(\ref{eq:5}) by substituting $|x_i - x_j|$ with $|x_i + x_j|$ and Eq.~(\ref{eq:6}) by
substituting $|x_i + x_j|$ with $|x_i - x_j|$ thus favouring similar opinions in link deletion and opposite
opinions in local rewiring, respectively, the final networks shows \textit {dissortative mixing} instead.

Furthermore, though not reported in detail in this paper, we have also performed numerical simulations to
investigate the effect of $y$ (i.e. the proportion of focal closure events in Eq.~(\ref{eq:2})) on the final 
configuration of the system. We have performed calculations in networks of size $N = 400$, keeping
$\left<k_0\right> = 4 $ and $g = 100$ constant, and varying $y$ from 0 to 1. The main result is that the
clustering coefficient remains practically constant, except when $y \approx 1$, where it decays. The average
shortest path is reduced when $y$ increases, which is to be expected, since focal closure offers the opening
of long jumps in the network in the same manner as disorder produces small world properties in a ring. Finally,
the average degree increases slightly with $y$, also explainable by the same token. More studies in this
direction will be made in the future.

\section{Mean field approximations}

In order to understand better the numerical results we analyze our model by using a mean field approach.
With this one can investigate, for instance, how rapidly the agents reach an irrevocable opinion, that is, how 
the mean number of decided (or undecided) agents behaves as a function of time in early stages of the 
dynamics, or when $t \approx 0$. If we take $h_i = 0$ as in our numerical calculations,
Eq.~(\ref{eq:2}) is symmetric with respect to the sign of $x_i$. In this equation the term 
$\alpha_i f_0(\{x_j\})$ is likely to be very small for two reasons:
\begin{enumerate}
\item for $t \approx 0$ the term $f_0(\{x_j\}) \sim 0$, because the distribution of $x_j$ is symmetric
around zero, and
\item $\alpha_i$ is a random parameter centered at 0, so when averaged over $i$ it turns out to be very small.
\end{enumerate}
Then one could write \[\dot{x}_i = x_i \sum_{j\in m_1(i)} x_j = \gamma_i x_i,\] which means that $x_i$ decays 
or grows exponentially at a rate given by the magnitude of $\gamma_i$ that varies locally. Without loss of 
generality we consider those $x_i$ that go to +1. Of all the agents with $x_i > 0$ only the ones with 
$\gamma_i > 0$ will reach eventually the maximum value 1, and this number is of the order of $N/4$.

For early times, the number of agents with $x_i = 1$, denoted by $N_1$, is small and the probability of
having a system of $N_1$ agents in a ``bath'' of $N$ should be of the canonical form
$P_1(N_1) = e^{-\mu' N_1} / Z'$, where the ``chemical potential" should fix the average number
$\left<N_1\right>$. Since the probability for all the states with a given value 
of $x_i$ is of the same form, and $N = N_0 + N_1$, the probability of having $N_0$ states with $x \ne 1$ 
is $P_0(N_0) = e^{-\mu N_0} / Z$, where
\begin{equation}
\label{eq:8}
\begin{split}
Z &= \sum_{N_0=0}^{N}e^{-N_0\mu} = \frac{1-e^{-\mu (N+1)}}{1-e^{-\mu}} \\
&= e^{\frac{-N\mu}{2}} \frac{\sinh\left( \frac{(N + 1)\mu}{2}\right)}{\sinh(\frac{\mu}{2})}.
\end{split} 
\end{equation}

Now, the average fraction of undecided agents $\left<n_0\right> = \left<N_0\right> / N$ at a given time can be 
calculated as
\begin{equation}
\label{eq:9}
\left<n_0\right> = \frac{1}{Z} \sum_{N_0=0}^{N} \frac{N_0}{N} e^{-\mu N_0} = -\frac{1}{N} \frac{\partial \ln Z}{\partial \mu}
\end{equation}

\begin{figure}[!ht]
\begin{center}
\epsfxsize = \columnwidth \epsffile{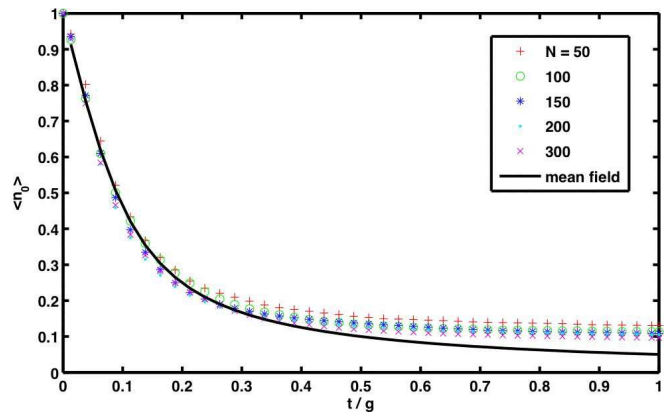} 
\caption{Average fraction of undecided agents $\left<n_0\right>$ as a function of time, in units of $g$, for
various networks of different size, where each point is an average over 1000 realizations. The mean field
prediction fitted with Eq.~(\ref{eq:10}) is shown in black.}
\label{fig:4}
\end{center}
\end{figure}

The quantity $\mu$ is difficult to calculate even with mean field approach so we will consider
it as an adjustable parameter. Here we assume that $\mu$ is of the form $t / \tau$, where $\tau$
is a large number. Taking into account both values of $\text{sign}(x_i^0)$ one finally gets
\begin{equation}
\label{eq:10}
\begin{split}
\left<n_0\right> &= 1 - \frac{N+1}{N} \coth\left( \frac{N+1}{2} \frac{t}{\tau}\right) + \frac{1}{N} \coth\left( \frac{t}{2\tau}\right) \\
 &= 1 - B_N \left(\frac{t}{\tau}\right),
\end{split}
\end{equation}
where $B_N (t / {\tau})$ is the Brillouin function found also for magnetic systems. In Fig.~\ref{fig:4}
we show the comparison of the mean field prediction of Eq.~\ref{eq:10} with numerical calculations performed in
networks of different sizes, and with a fixed value of $g = 10^5$. The parameter $\tau$ was fitted with a least
squares technique and has a value of approximately $gN / 40$, which for $N \ge 50$ is indistinguishable from
the Langevin classical limit, independently of $N$. Then the characteristic time taken by an agent to reach
its definite opinion is $g / 20$, so almost all the agents have reached their definite opinions before rewiring.

Notice that this result does not depend on the size of the network, and that the fitting is very good for
short times, but starts to deviate significantly and systematically for longer times. In the numerical
calculations more agents remain undecided than what is predicted by our mean field approximation. This is
to be expected, since we have not taken into account the effects of frustration due to changes of opinion
produced by the distribution of the personal attitude parameter $\alpha$. Indeed, the difference between the
mean field prediction and the numerical value can be used as a measure of the amount of frustration, which
can become important in subsequent rewirings.

So far we have analyzed the fast dynamics of the state variable but have not considered the rewiring processes.
An interesting thing to do is to compare the deviations of our model results from the known properties of a
random network, for instance 
\begin{equation}
\label{eq:11}
\left<k\right> = \left<C\right> N.
\end{equation}
Notice that in our model $\left<k\right>$ should be constant to first order, since the network changes are
such that every time a link is deleted another one must be created. Therefore, a change of degree $\Delta k_i$
is due to simultaneous actions of both agents connected by a link in the parallel rewiring scheme.

In each rewiring event every agent deletes and creates the same number of links, so the agent's degree does
not change. However, if a neighbor decides to delete the same link and create a different one, then two links
are created but only one deleted resulting in the average degree of the network to increase. Similarly, if
two agents independently decide to create a link between them and both have deleted different links, the 
average degree decreases. This can be expressed as follows
 \begin{equation}
\label{eq:12}
\Delta k_i = \Delta k_i^+ - \Delta k_i^-,
\end{equation}
where the first term stands for the correlated processes associated to link deletion increasing the degree
and the second term for the processes associated to link creation decreasing the degree. Of the $k_i$ first
neighbors of node $i$ that could be deleted, only a fraction $f_i^+$ of them will produce a correlated
increase of the average degree. Then one could write $\Delta k_i^+ = f_i^+ k_i$.

In our numerical simulations we have considered only triadic closure mechanism as the process of creating
links. Thus, the second term in Eq.~(\ref{eq:12}) should be proportional to the number of second neighbors 
$n_i^{(2)}$, or $\Delta k_i^- = f_i^- n_i^{(2)}$. Now, only a fraction of the $(k_j-1)$ second neighbors of $i$
connected to first neighbor $j$ will produce a correlated decrease of the average degree. Let us denote this
number by $(f_j^- k_j-1)$. However, this holds only in a tree, where all the second neighbors are different,
but in a general network there should be cyclic closures, like triangles ($n^\bigtriangleup$) and squares
($n^\square$). Then one may write
\begin{equation}
\label{eq:13}
n_i^{(2)} = \sum_{j\in m_1(i)} (f_j^- k_j - 1) - 2n_i^\bigtriangleup - n_i^\square,
\end{equation}
where the summation runs over the first neighbors of $i$. The factor of 2 in the second term is due to the
fact that there are two triangles associated with a single triad. The exact count of links, according to
Eq.~(\ref{eq:12}) is
\begin{equation}
\label{eq:14}
\begin{split}
& \left< \Delta k \right> = \\
& \frac{1}{N} \sum_{i=1}^N \left[ f_i^+ k_i - f_i^- \left( \sum_{j\in m_1(i)} (f_j^- k_j - 1) - 2n_i^\bigtriangleup - n_i^\square \right) \right].
\end{split}
\end{equation}
The number of triangles is related to the clustering coefficient as
\begin{equation}
\label{eq:15}
C_i = \frac{2n_i^\bigtriangleup}{k_i (k_i - 1)}.
\end{equation}

So far we have not made any approximations. The factors $f_i^+$ and $f_i^-$ are time dependent through the 
dynamics of the state variables $x_i$. It is clear that these factors are non-zero only if the two sites
involved in the same rewiring process are still undecided, that is, the probability of having one of these 
correlated rewirings should be proportional to $\left<n_0\right>^2$, and can be calculated numerically by
keeping track of the rewiring matrices $P_{ij}$ and $Q_{ij}$ defined as
\begin{equation}
\label{eq:16}
P_{ij} = 
\begin{cases}
-1 & \textrm{ if   } j \in N_i^{(1)} | p_{ij} > 0 \\
0 & \textrm{ othewise   } 
\end{cases},
\end{equation}
and
\begin{equation}
\label{eq:17}
Q_{ij} = 
\begin{cases}
1 & \textrm{ if   } j \in N_i^{(2)} | q_{ij} > 0 \\
0 & \textrm{ othewise   } 
\end{cases},
\end{equation}
where $N_i^{(1)}$ and $N_i^{(2)}$ are subsets of the first and second neighbors of agent $i$, such that their
cardinalities are the same because rewiring is the only operation allowed (see Eqs.~(\ref{eq:5})
and (\ref{eq:6})). Then it is clear that
\begin{eqnarray}
\label{eq:18}
f_i^+ = \frac{(P^2)_{ii}}{(P^2)_{ii} + (Q^2)_{ii}}, & \textrm{ and  } & f_i^- = \frac{(Q^2)_{ii}}{(P^2)_{ii} + (Q^2)_{ii}}. \nonumber \\
\end{eqnarray}

The normalization factor comes from the fact that only operations taking place by both agents $i$ and $j$ 
simultaneously contribute to the count in $i$. This is proportional to 
$\sum_j (P_{ij} + Q_{ij})^2$ where $P_{ij} Q_{ji} = 0$. Then $f_i^- = 1-f_i^+ = f_i$. One can neglect the term 
$n_i^\square$ in Eq.~(\ref{eq:14}) because the number of squares should be small as compared with the 
number of triangles. Also, we may assume that $f_i$ is very similar in all sites and near to 1/2. In fact one 
can trace the mean value of $f_i$ as a function of time in the numerical calculations and see that it converges
very rapidly to the value 1/2. Performing the average over $i$ and assuming the degree, the fraction of nodes,
and the clustering coefficient statistically independent we obtain
\begin{equation}
\label{eq:19}
\Delta k = \left< n_0\right>^2 \left[(1 - f)k - fk(fk - 1) + fCk(k - 1)\right],
\end{equation}
where we have suppressed the brackets indicating averages and dropped the subindex in all the quantities.
Therefore, we find a fixed point when
\begin{equation}
\label{eq:20}
k = \frac{1 - fC}{f(f - C)}.
\end{equation}

\begin{figure}[!ht]
\begin{center}
\epsfxsize = \columnwidth \epsffile{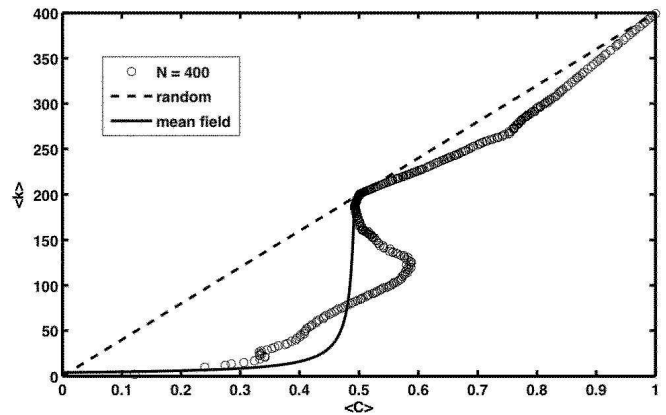} 
\caption{The average final degree as a function of the average final clustering coefficient obtained from 
simulations in networks of size $N = 400$ and $g = 100$, and several initial degrees $\left<k_0\right>$ 
(open circles). The prediction by Eq.~(\ref{eq:11}) for a random network is shown as a dashed line, and the 
mean field result of Eq.~(\ref{eq:20}) is shown as a continuous line.}
\label{fig:5}
\end{center}
\end{figure}

In Fig.~\ref{fig:5} we plot the result of Eq.~(\ref{eq:20}) with $f = 1/2$ (continuous line) and compare
it with the numerical results (open circles) and the linear dependence predicted by Eq.~(\ref{eq:11}) for
a random network (dashed line).  Note that for any given average degree $k$ the clustering coefficient for
the mean field and numerical results is larger than for a random network, and for $ k = N/2$ all the
clustering coefficients meet at value 1/2. Also we would like to point out that although in the mean field
approximation we have taken only a part of the second order effects into account, it behaves qualitatively
the same as the numerical result. 

\section{Concluding remarks}
\label{con}

In this paper we present a general approach to the problem of opinion formation on an adaptive network,
which can be considered as a basic problem of social network formation and communication. The dynamical
equation Eq.~(\ref{eq:1}) and its modification Eq.~(\ref{eq:2}) contain several aspects of this
problem, which to our knowledge have not been considered or addressed before. First, the role of
general opinion of all the agents in the system representing the overall mood sensed by each individual
with his/her own attitude towards it. Second, the effect of the media as a kind of external field term, not as
a constant field as suggested before (see, e.g., \cite{HKS2001}) but as an agent dependent reaction term
corresponding to a personal attitude towards the ``manipulation'' attempts by the media. This makes the
system reminiscent to a random field model. Third, in our model the two time scales of the basic dynamical
processes are clearly separated and deterministically controlled, unlike in other studies where an additional
parameter like bounded confidence controls the range within which different opinions act on each
other \cite{KBJPA2008}, thus affecting the time scale at which opinions change. We note that also in our
model the bounded confidence parameter could be included, but we chose not to do so in the interest of
not increasing the number of adjustable parameters in the model. Fourth, in our model the rewiring has
been done deterministically favoring the triadic closure mechanism rather than random rewiring mechanisms
used in other studies \cite{NKB2008,KB2008,KBJPA2008,mandra}. In our model both the basic link-rewiring
mechanisms of network sociology, namely focal and triadic closure, have been included but in this paper we
have mainly focused on the effects of the latter rewiring mechanism. 

In this study rather than exploring the entire richness of our model, we have concentrated on the effect of the
separation of time scales for comparison with previous results. We have shown that the important feature of
coevolution is the separation of the two basic time scales, namely, the rapid dynamics of the state variable,
and the slow dynamics of the network rewiring. The only parameter of our model has been $g$, serving as
a measure of this separation.

As one of the key results we have found that for intermediate values of the time scale variable $g$ the
network turns out to organize in well-connected small communities of agents with the same opinion. This is
in accordance with the earlier study \cite{KBJPA2008}, where it was found that communities could form when
the tolerance or bounded confidence parameter was varied, which could be understood to be due this parameter
changing the effective time scale for rewiring. In addition we have studied further the role of the difference
in time scales for fast and slow dynamics analytically by devising a mean field treatment. Using random
network as a reference, we have found that the mean field results compare quite well with the numerical
results and both of them differ significantly from the random network results.

In the future we plan to investigate our model systematically and in detail for the effects of personal
attitude parameter $\alpha_i$ and the personal random field term of an agent $h_i$, as well as the relative
importance parameter $y$ between local rewiring with triadic closure and global rewiring with focal closure
mechanisms. Furthermore, another interesting point to investigate in our model would be the relative importance
of short and long range interactions. 

As a final remark we believe that our fully dynamical approach and the kind of coevolving network model could
be applied to various other situations, not only in social networks but in other fields, like symbiotic
relations between two species sharing an ecological environment. Some of these applications are currently a
matter of further explorations.

\acknowledgements

K.~K. acknowledges the Academy of Finland, the Finnish Center of Excellence 
programme 2006 - 2011, proj. 129670. J.~K. acknowledges partial support from Helsinki University of Technology
visiting professor programme and OTKA K60456 grant. K.~K. and R.~A.~B. want to acknowledge financial support
from Conacyt through project 79641. R.~A.~B. is grateful to Helsinki University of Technology for an adjunct
professorship and for hospitality in the Centre of Excellence in Computational Complex Systems Research,
where most of this work has been done.

\end{document}